\documentclass[prd,superscriptaddress,twocolumn,floatfix]{revtex4}

\usepackage{natbib} 
\usepackage{bm}

\usepackage{graphicx}
\usepackage{amsmath}
\usepackage{amssymb}
\usepackage{hyperref}

\newcommand{\LCDM}{$\Lambda\mathrm{CDM}$}

\newcommand{\tcm}{21\,cm}

\begin{document}

\title{Near term measurements with 21\,cm intensity mapping: 
neutral hydrogen fraction and BAO at $z<2$}

\author{Kiyoshi Wesley Masui}
\email{kiyo@cita.utoronto.ca}
\affiliation{Canadian Institute for Theoretical Astrophysics, 60 St.~George Street
Toronto, Ontario, M5S 3H8, Canada}
\affiliation{Department of Physics, University of Toronto, 60 St.~George Street
Toronto, Ontario, M5S 1A7, Canada}

\author{Patrick McDonald}
\email{pmcdonal@cita.utoronto.ca}
\affiliation{Canadian Institute for Theoretical Astrophysics, 60 St.~George Street
Toronto, Ontario, M5S 3H8, Canada}

\author{Ue-Li Pen}
\email{pen@cita.utoronto.ca}
\affiliation{Canadian Institute for Theoretical Astrophysics, 
60 St.~George Street
Toronto, Ontario, M5S 3H8, Canada}

\date{May 2, 2010}

\begin{abstract}

It is shown that \tcm{} intensity mapping could be used in the
near term to make cosmologically useful measurements.
Large scale structure could be detected using existing radio
telescopes, or using prototypes for dedicated redshift survey
telescopes.  This would provide a measure of the mean neutral
hydrogen density, using redshift space 
distortions to break the degeneracy with the linear bias.
We find that with only
200 hours of observing time on the Green Bank Telescope,
the neutral hydrogen density could be measured to 25\% 
precision at redshift $0.54<z<1.09$.  This compares favourably to current
measurements, uses independent techniques, and
would settle the controversy over an important parameter which impacts
galaxy formation studies.
In addition, a 4000 hour survey would allow for the detection of
baryon acoustic oscillations,
giving a cosmological distance measure at 3.5\% precision.  These 
observation time requirements could be greatly reduced with the construction
of multiple pixel receivers.
Similar results are 
possible using prototypes for dedicated cylindrical 
telescopes on month time scales, or SKA pathfinder aperture
arrays on day time scales.  Such measurements
promise to improve our understanding of these quantities while
beating a path for future generations of 
hydrogen surveys.

\end{abstract}

\maketitle

\section{Introduction}
\label{s:intro}
An upcoming class of experiments propose the observation of the 
\tcm{} spectral line over large volumes, detecting large scale
structure in three dimensions \cite{Peterson:2009ka}. 
This method is sensitive to
a redshift range of $z \gtrsim 1$, which is
observationally difficult for
optical experiments, due to a dearth of spectral lines in
the atmospheric transparency window.
Such experiments require only limited resolution to resolve the structures of
primary
cosmological interest, above the non-linear scale.  At $z \approx 1$ this
corresponds to tenths of degrees.  There is no need to detect 
individual galaxies, and in general each pixel will contain many.
This process is referred to as \emph{\tcm{} intensity mapping}.
A first detection of large scale structure in the \tcm{} intensity
field was reported in 2008 \cite{arXiv:0802.3239}.

Intensity mapping is sensitive to the large scale power spectra in
both the transverse and longitudinal directions.  From this signal,
signatures such as the baryon acoustic oscillations
(BAO), weak lensing and redshift space distortions (RSD) can be detected
and used to gain cosmological insight.

To perform such a survey dedicated cylindrical radio
telescopes have been proposed \cite{Peterson:2006bx,Seo:2009fq},
which could map a large fraction of
the sky over a wide redshift range and on timescales of several years.
These experiments are economical since their
low resolution requirements imply limited size and they have 
no moving parts.  It has been shown that BAO detections from
\tcm{} intensity mapping are powerful probes of 
dark energy, comparing favourably with Dark Energy Task Force
Stage IV projects within the figure of merit framework
\cite{Chang:2007xk,Albrecht:2006um}.  Additionally, in
\cite{Masui:2009cj} it was shown
that such experiments could tightly constrain theories of
modified gravity, making extensive use of weak lensing information.
The $f(R)$ theory, for instance, could be constrained nearly
to the point where the chameleon mechanism masks any deviations from 
standard gravity before they first appear.

While such experiments will be very powerful probes of the
Universe, it is useful to explore how \tcm{} intensity mapping could
be employed in the short term.  Here we discuss surveys that
could be performed using existing radio telescopes, where the Green Bank 
Telescope will be used as an example.  Prototypes for the
above mentioned cylindrical telescopes (for which some
infrastructure already exists) are also considered.
Finally, the Square Kilometre Array Design
Studies (SKADS) focused on developing aperture array 
technology for the Square Kilometre Array \cite{skads}.  
Motivated by the proposed
Aperture Array Astronomical Imaging Verification (A3IV)
project \cite{ardenne, aavp}, we
consider pathfinders for such arrays.
These aperture arrays could
share many characteristics in common with cylindrical
telescope prototypes, but would be capable of much greater
survey speed.

While these limited resources would not have nearly the
statistical power required to detect effects like 
weak lensing, a detection of the RSD would be possible.  This
would give a measure of the mean density of neutral hydrogen
in the Universe.
This has been an important and controversial parameter
in galaxy formation studies and a precise measurement would be invaluable in 
this field \cite{arXiv:0902.4717}.
In addition BAO are considered, a detection of which would
yield cosmologically useful information about the Universe's expansion
history.

In this paper we first describe the RSD
and BAO and the information
that can be achieved with their detection.  We then present forecasts
for \tcm{} redshift surveys as a function of telescope time, followed
by a brief discussion of these results.

We assume a fiducial \LCDM{} cosmology with parameters: 
$\Omega_m = 0.24$, $\Omega_b = 0.042$, $\Omega_\Lambda = 0.76$, 
$h = 0.73$, $n_s=0.95$ and $\log_{10}A_s = -8.64$; where these
represent the matter, baryon and vacuum energy densities 
(as fractions of the critical density), the dimensionless Hubble
constant, spectral index and logarithm of the amplitude of scalar
perturbations.

\section{Redshift Space Distortions}

In spectroscopic surveys, radial distances are given by redshifts.
However redshift does not map
directly onto distance as matter also has peculiar
velocities, which Doppler shifts incoming photons.  On large scales,
these velocities are coherent and result in additional apparent 
clustering of matter in redshift space.
In linear theory, the net effect is an enhancement
of power for Fourier modes with wave vectors along the line of sight
\cite{1987MNRAS.227....1K},
\begin{equation}
	P^{\rm s}_X(\vec{k},z) = b^2 \left[1+\beta(z) \mu_{\hat{k}}^2 \right]^2 P(k,z).
	\label{e:RSD:P_s}
\end{equation}
Here $P^s_X$ is the power spectrum of tracer $X$ (assumed to be linearly
biased) as observed in redshift space, $P$ is the matter power
spectrum and 
$\mu_{\hat{k}} = \hat{k}\cdot\hat{z}$.  The bias, $b$, quantifies
the degree to which the density perturbation of the tracer follows
the density perturbation of the underlying dark matter. The redshift
space distortion parameter $\beta$ is equal to $f/b$ in linear
theory, where
$f$ is the dimensionless linear growth rate. To a good approximation
$f(z) = \Omega_m(z)^{0.55}$. 

We define the signal neutral hydrogen power spectrum as
\begin{equation}
	\tilde{P}_{\rm HI}(\vec{k},z) \equiv x_{\rm HI}^2 P^{\rm s}_{\rm HI}(\vec{k},z),
	\label{e:RSD:Psignal}
\end{equation}
where $x_{\rm HI}(z)$ is the mean fraction of hydrogen that is neutral.
It is
the parameter $x_{\rm HI}$ that we wish to measure.
The above definition is useful because it defines the quantity that
is most directly measured in a \tcm{} redshift survey.  
Note that in discussions of the more standard galaxy redshift surveys one
usually takes
a measurement of the mean density for granted; however, \tcm{} intensity 
mapping will make differential measurements and
will not measure the mean signal directly.  One cannot divide out the mean to 
define a measurement of fluctuations around the mean.
Using the
fact that at late times, on large scales, structure undergoes scale 
independent growth, the above quantity can be written as
\begin{equation}
	\tilde{P}_{\rm HI}(\vec{k},z) = b^2 x_{\rm HI}^2 \left(\frac{g(z) a}{a_0}\right)^2
		\left[1+\beta \mu_{\hat{k}}^2 \right]^2 P(k,z_0),
	\label{e:RSD:PsHIP0}
\end{equation}
where subscript $0$ refers to some early time well after recombination 
and $g$ is the growth factor (relative to an Einstein-de Sitter Universe).

Because the power spectrum at early times, $P(k,z_0)$, can be inferred from
the Cosmic Microwave Background (to good enough accuracy for our purposes in 
this paper), the observed power spectrum can
be parameterized by just two redshift dependent numbers, $\beta$
and the combination of scale independent prefactors in Equation
\ref{e:RSD:PsHIP0},
\begin{equation}
	A_H \equiv b^2 x_{\rm HI}^2 g^2 .
	\label{e:RSD:AH}
\end{equation}
It is these two parameters that can be determined from a \tcm{}
redshift survey using the RSD.  In general $A_H$ will be better
measured than
$\beta$ and does not significantly contribute to the uncertainty
in $x_{\rm HI}$.

The factors $f(z)$ and $g(z)$ depend on the expansion history.
If one allows for a general expansion history (for instance,
the WMAP allowed CDM model with arbitrary curvature and equation 
of state, OWCDM) these factors are poorly determined since there
is currently little data that directly probes the expansion
in this era.  However, if one is willing to assume a flat
\LCDM{} expansion history, then the current uncertainty in the
late time expansion is attenuated at the redshifts of interest
(since dark energy is sub-dominant to matter at $z=1$)
and uncertainties in theses parameters can be ignored.

As such, a measurement of $\beta$ gives a measurement of the
bias $b$ which in turn gives a measurement of $x_{HI}$.
We have
\begin{equation}
	\frac{\Delta x_{\rm HI}}{x_{\rm HI}} \approx \frac{\Delta \beta}{\beta}
	\qquad \text{(\LCDM{} assumed)}.
	\label{e:RSD:dx}
\end{equation}

If one is not willing to assume an expansion history, it is a
simple matter to propagate the corresponding uncertainties in
the expansion parameters.

To estimate errors on $\beta$, we assume that the \emph{primordial} power
spectrum is essentially known from the cosmic microwave background,
and we fix all parameters except for the amplitude $A_s$, spectral
index $n_s$, and running of the spectral index $\alpha_s$.
The \emph{observable} power
spectrum is then parameterized by $\beta$, $A_s$, $n_s$ and
$\alpha_s$.  We then
use the Fisher matrix formalism to determine how precisely $\beta$
can be measured from the 21\,cm survey, marginalizing over the other
three parameters and using no other information.
The parameter $A_s$ is used as a stand-in for the other parameters
that affect the overall amplitude of the power spectrum: the bias
and $x_{\rm HI}$.  Its marginalization is 
critical to account for the fact that we have no a priori
information about these parameters.
The spectral index marginalization is not strictly necessary but 
allows for some degradation
due to concern about scale dependence of bias. We have restricted the 
numerator of the Fisher matrix to include only the linear theory power as
suppressed by the non-linear BAO erasure kernels of 
\cite{2007ApJ...665...14S} (all the
linear power is included though, not BAO only),
so non-linearity should not be 
a significant issue at the level of precision discussed here.
This treatment of the 
non-linearity cutoff is also motivated by the propagator work of 
\cite{2006PhRvD..73f3520C,2006PhRvD..73f3519C,2008PhRvD..77b3533C}.

\section{Baryon Acoustic Oscillations}

Acoustic oscillations in the primordial photon-baryon plasma
have ubiquitously left a distinctive imprint in the distribution of
matter in the Universe today.  This process is understood from
first principles and gives a clean length scale in the Universe's 
large scale structure, largely free of systematic uncertainties, and 
calibrations.  This can be used to measure the global cosmological 
expansion history through the angular diameter distance vs 
redshift relation. The detailed expansion will 
differ between a pure cosmological constant and the various other 
cosmological models.

\begin{figure}
	\centerline{\includegraphics[scale=1]{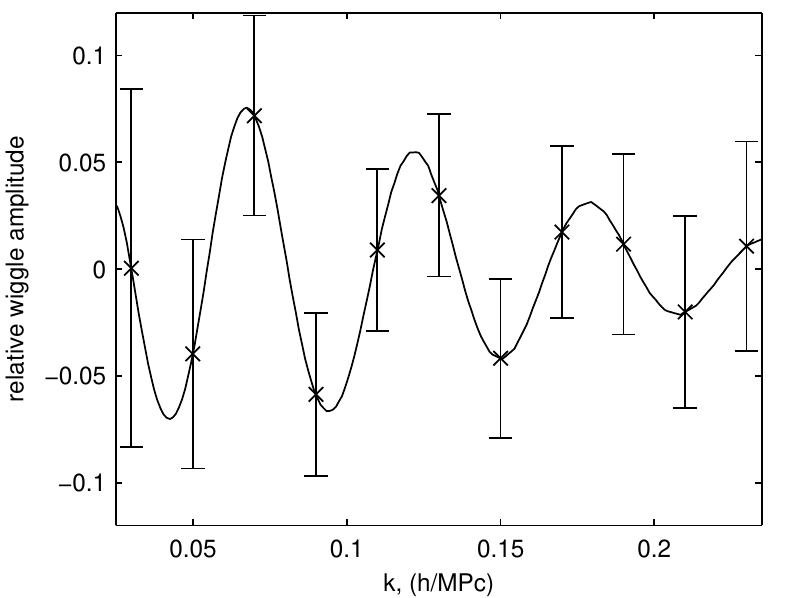}}
	\caption{\label{f:BAO:wiggles} 
	Baryon acoustic oscillations averaged over all directions.
	To show the BAO we plot the ratio of the full matter power spectrum to 
the wiggle-free power spectrum of \cite{1998ApJ...496..605E}.
The error bars represent projections
of the sensitivity possible with 4000\,hours observing time on GBT
at $0.54 < z < 1.09$.
}
\end{figure}

We use essentially the method of \cite{2007ApJ...665...14S} for estimating
distance errors obtainable from a BAO measurement, including 50\% 
reconstruction of nonlinear degradation of the BAO feature (although this
is unimportant since experiments considered here have low resolution).
The BAO feature is isolated by dividing the total power spectrum 
\cite{1998ApJ...496..605E} by the wiggle-free power spectrum  
and subtracting unity,
as illustrated in Figure \ref{f:BAO:wiggles}.
The wiggles are then parameterized by an overall amplitude, and a
length scale dilation (here $A_{w}$ and $D$ respectively), which
control the vertical and horizontal stretch of the 
theoretical curve
shown in Figure \ref{f:BAO:wiggles}. 
Our errors on $A_{w}$ come from a straightforward extension of the 
\cite{2007ApJ...665...14S} method for estimating BAO errors. In addition
to the BAO distance scale as a free parameter in our Fisher matrix, we include 
$A_w$ as a free
parameter. This is similar to what one sometimes tries to do by including the
baryon/dark matter density ratio as a parameter, but more straightforward to 
interpret.

The ability to measure $A_w$ (which is 
zero in the absence of the BAO) represents the ability to detect the presence
these wiggles.  A measurement of $D$ allows one to associate
a comoving distance to length scales on the sky.  This gives a 
measurement of the angular diameter distance ($d_A$) for detections
in the transverse direction, and
the Hubble parameter ($H$) if the wiggles are detected in the longitudinal
direction.

\section{Forecasts}

We present forecasts for the Green Bank
Telescope and prototypes for two classes of telescope:
cylindrical telescopes and SKA aperture arrays.
The signal available for 21\,cm experiments is 
proportional to the neutral hydrogen fraction and bias.
For estimating telescope sensitivity
we assume that the product of the bias and the neutral 
hydrogen density $\Omega_{\rm HI}b = 0.0004$ today\cite{Chang:2007xk},
and that the neutral hydrogen fraction and bias do not evolve.
These assumptions only affect the sensitivity of the telescopes and
not the translation from uncertainty on $P^s_{\rm HI}$ to the
uncertainty on $x_{\rm HI}$.
Also, as in galaxy
surveys, there is expected to stochastic shot noise component and we
assume Poisson noise with an effective
object number density $\bar{n}=0.03$ per cubic $h^{-1}\rm{Mpc}$.
Note that stochastic noise at this level is
negligible, as should be the case in practice.

The 21\,cm intensity mapping technique is expected to be
complicated by a variety of contaminating effects.  These
include diffuse foregrounds (predominantly galactic
synchrotron), radio frequency interference and bright point sources.
The degree to which these contaminants will limit future surveys has
yet to be quantified, and here we simply ignore them.  As such these
forecasts are theoretical idealizations.  Methods for dealing with
these contaminants are discussed in \cite{Chang:2007xk}.

\begin{figure}
	\centerline{\includegraphics[scale=1]{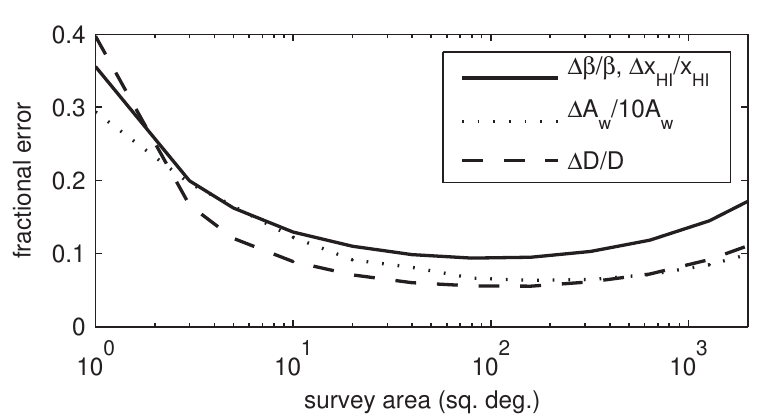}}
	\caption{\label{f:results:gbtoptarea} Ability of GBT to measure
	the BAO and redshift space distortions as a function of 
	survey area at fixed observing time.  Presented survey is
	between $z = 0.54$ and $z = 1.09$ and observing time is
	1440\,hours.  A factor of 10 has been removed from the $A_w$ 
    curve.}
\end{figure}

\begin{figure}
	\centerline{\includegraphics[scale=1]{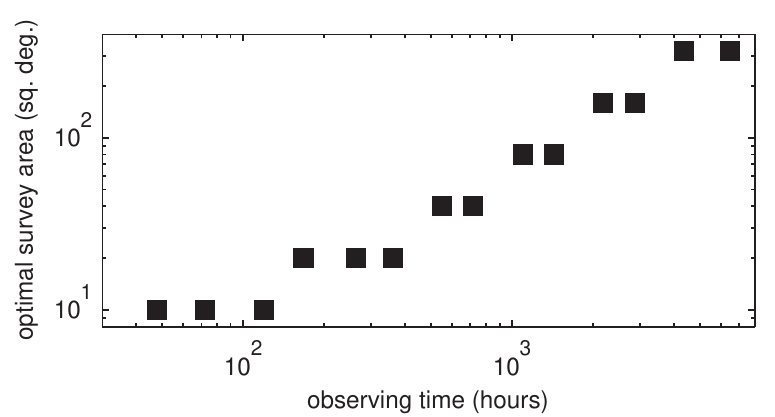}}
	\caption{\label{f:results:gbtarea} Roughly optimized survey area
	as a function of telescope time on GBT.  Redshift range is
    between $z = 0.54$ and $z = 1.09$.}
\end{figure}

The Green Bank Telescope is a 100\,m diameter circular telescope
with a system temperature of 25\,K.
It has interchangeable single pixel receivers at the frequencies of
interest with bandwidths of approximately 200\,MHz.  For extended 
surveys, multiple pixel receivers could be implemented.  The construction
of a four pixel receiver is within reason and would reduce the 
required telescope time by a factor of four.
In planning  a survey on GBT, it is important to
choose an appropriate survey area.  As illustrated in Figure
\ref{f:results:gbtoptarea}, at fixed observing time, there
is a survey area that best measures the desired parameters.
For all results the survey
area has been roughly optimized for the
quantity being measured.  The optimized areas are shown in Figure
\ref{f:results:gbtarea}.  Results are essentially insensitive to this 
area within a factor of 2 of the optimum.

Prototyping for dedicated cylindrical telescopes is in its
early stages.  We present forecasts for a hypothetical 
not-too-far-off telescope, composed of two cylinders.
The total array measures 40\,m$\times$40\,m with 300 dipole receivers 
with 200\,MHz
bandwidth, and a system temperature of 100\,K.
Such telescopes have no moving parts
and point solely by the rotation of the earth.  As such, the
area of the survey is set by latitude, receiver spacing, and obstruction
by foregrounds; we assume
15\,000 square degrees.  The survey area scales with the number of 
receivers leaving the noise per pixel unchanged.  Thus the squared errors on 
measured quantities scale inversely with the number
of receivers.
Note that this is a slightly different scaling than the
area optimized case of GBT, where it is the time axis that is 
scaled by the number of receivers.
However, in the area optimized case variances also scale inversely
with time as area optimization effectively fixes the noise per pixel.
As such our results for GBT follow both scalings.

The forecasts for cylindrical telescopes are also
applicable to pathfinder aperture arrays \cite{skads}.
Where-as cylinders
use optics to form a beam in one dimension and interferometry
in the other, aperture arrays directly sample the incoming
radio waves without optics.  Interferometry is used in both
dimensions, forming a two dimensional array of beams, instead
of a two pixel wide strip for a two cylinder telescope.
In principle it would be possible for an
aperture array to monitor essentially the whole sky
simultaneously, and thus form thousands of beams.
In practise digitizing every antenna is costly and many
antennas must be added in analogue in such a way that some
beams are preserved but many are cancelled.
We consider a compact aperture array
that has the same area and resolution as the cylinder considered
here. This is almost
identical in scale as the proposed A3IV \cite{ardenne}.  We
assume preliminary A3IV specifications, with 700 effective
receivers, 300\,MHz bandwidth, and 50\,K system temperature
\cite{ger}.  The aperture array could thus perform the same
survey as the cylinder but a factor of
$(300\,\mathrm{MHz}/200\,\mathrm{MHz})
(700/300)(100\,\mathrm{K}/50\,\mathrm{K})^2=14$ faster.  Note
however, that for aperture arrays there is added freedom in
which beams are preserved when antennas are added.  It would be
thus possible to optimize the area of the survey as in the
GBT case.

\begin{figure}
	\centerline{\includegraphics[scale=1]{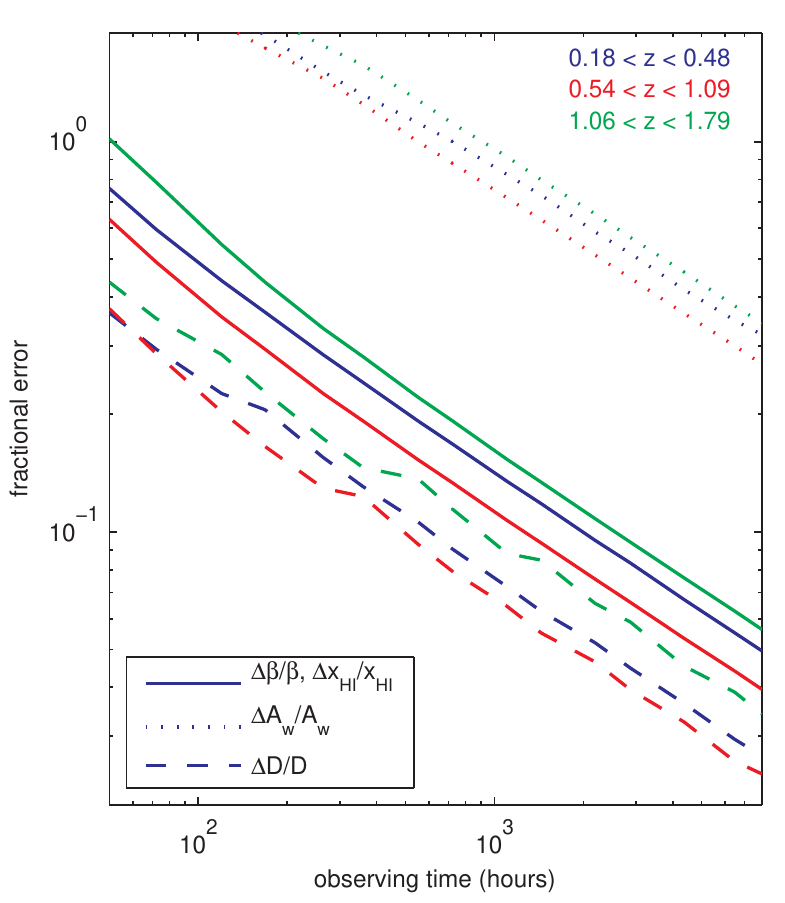}}
	\caption{\label{f:results:gbt} (Color online) Forecasts for fractional error
	on redshift space distortion and baryon acoustic oscillation
	parameters for intensity mapping surveys on the Green
	Bank Telescope (GBT).  Frequency bins are approximately 
	200\,MHz wide
	and correspond to available GBT receivers.  Uncertainties on 
	$D$ should not be trusted unless the uncertainty on $A_w$ is less
	than 50\% (see text).}
\end{figure}

\begin{figure}
	\centerline{\includegraphics[scale=1]{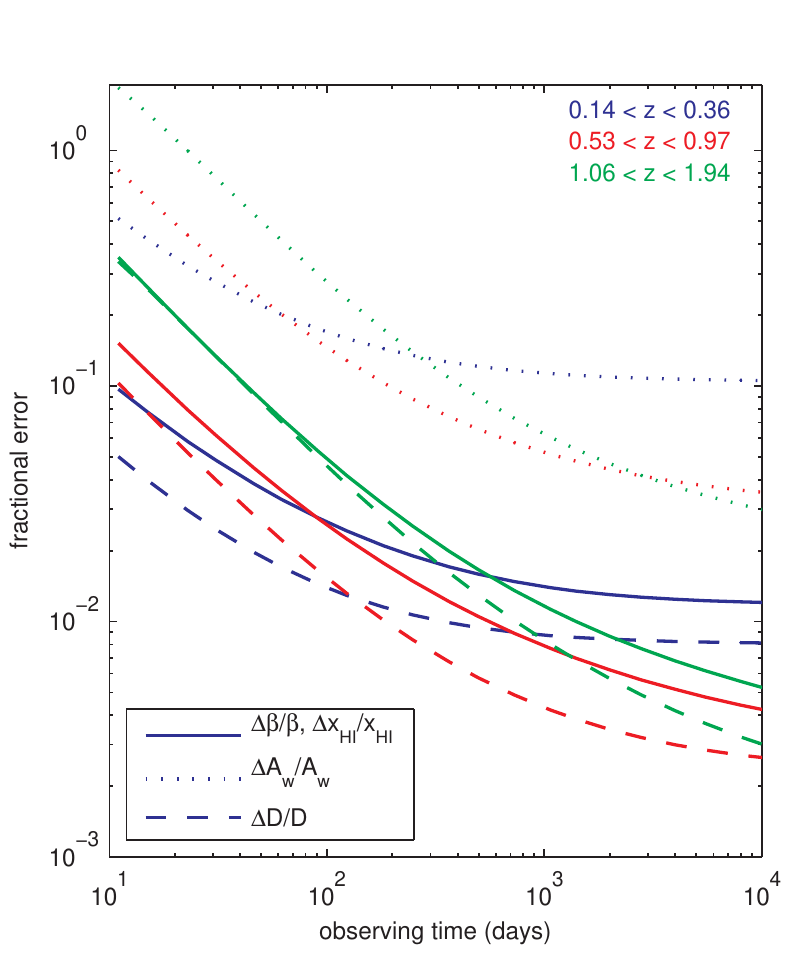}}
	\caption{\label{f:results:cyl} (Color online) Forecasts for fractional error
	on redshift space distortion and baryon acoustic oscillation
	parameters for intensity mapping surveys on a prototype cylindrical
	telescope.  Frequency bins are 200\,MHz wide corresponding to the
	capacity of the correlators which will likely be
    available. These result also apply to the aperture
    telescope but with the observing time reduced by a factor
    of 14.
	Uncertainties on 
	$D$ should not be trusted unless the uncertainty on $A_w$ is less
	than 50\% (see text).  Observing time does not account for
	lost time due to foreground obstruction.}
\end{figure}

Figures \ref{f:results:gbt} and \ref{f:results:cyl} show
the obtainable fractional errors on RSD parameter $\beta$ and 
BAO parameters $A_w$, the amplitude of wiggles, and $D$ the
overall dilation factor.  $D$ is defined as a simultaneous
dilation in both the longitudinal and transverse directions, i.e. 
both $H$ and $d_A$ are proportional to $D$.
There is another ``skew'' parameter which trades one for the other,
such that the Hubble parameter and the angular diameter
distance can be independently determined, however, this
parameter is generally not as precisely measured
\cite{2008PhRvD..77l3540P}.  $D$ is the 
mode that contains
most of the information available in the BAO and
we marginalize over the skew parameter.  The marginalized error on 
$D$ is independent of the exact definition of the skew.
Fractional errors on $H$ 
and $d_A$ are of order twice the fractional error
on $D$.  

Referring to Figure \ref{f:BAO:wiggles} above, it can be seen 
intuitively that
the parameters $D$ and $A_w$ are weakly correlated.  Furthermore,
since the signal is linear in the parameter $A_w$, our linear
Fisher analysis applies even for large errors.  This however cannot
be said about $D$.  Indeed any uncertainty on $D$, that brings the
smallest scale peak we resolve more than $\sim \pi/2$ out of phase,
cannot be trusted.  It can be seen from Figure \ref{f:BAO:wiggles}
that this corresponds to a fractional error of order 10\%, 
which of course depends on resolution.
For this reason, we require that the uncertainty on 
$A_w$ must be at most 50\% (a 95\% confidence detection of the BAO)
before we have any faith in the uncertainty in $D$.  We note that
Fisher analysis is not the optimal tool for determining when
an effect can first be measured; however, it applies for any
cosmologically useful measurements of the BAO.

\section{Discussion}

Typically measurements of the neutral hydrogen density
at high redshift are made using damped Lyman-$\alpha$
(DLA) absorption lines; current 
measurement uncertainties being
at the 25\% statistical precision level in the $0.5<z<2$ redshift range
\cite{Prochaska:2005wy,Rao:2005ab}.
However, it has been argued that these measurements 
are biased high \cite{Prochaska:2008fp} rendering the quantity
effectively uncertain by a factor of 3.  
Hydrogen is the main baryonic component in
the universe and it becomes neutral after falling into
galaxies and becoming self shielded from ionizing radiation.
As such, the abundance of neutral hydrogen is linked to the
availability of fuel for star formation
\cite{1986ApJS...61..249W,1995ApJ...454...69P}.
Understanding how the neutral hydrogen evolves over cosmic
time is key to understanding the star formation history 
and feedback processes in galaxy formation
studies \cite{Wolfe:2005jd,Shen:2009zd}.
Additionally, the linear bias, and any scale dependence it
might have, gives valuable information about how the gas is
distributed.
Finally, these quantities are crucial
for estimating the sensitivity of future \tcm{} redshift surveys
since the signal is proportional to the product of the bias
and the mean neutral density \cite{Chang:2007xk}.  
  
We have shown that even with existing telescopes,
it is possible to use 21\,cm intensity mapping to make useful
measurements of large scale structure at high redshift.  As seen
in Figure \ref{f:results:gbt}, a 4\,$\sigma$ detection of redshift
space distortions could be made at $z\approx 0.8$ with only 200 hours of
telescope time at GBT.  This would provide a 25\% measurement of the neutral
hydrogen fraction in the Universe using methods independent of DLA
absorption lines.  A longer survey using 1000 hours of telescope time
could make $\sim$$12\%$ measurements.

Surveys extending to this level of precision become cosmologically
useful. With 4000 observing hours on GBT (1000 hours with a four pixel
receiver), the BAO overall distance 
scale could be
measured to 3.5\% precision at the same redshift.  This is approaching
the precision of the WiggleZ survey, which will make a $\sim$$2\%$
measurement of this scale over a similar redshift range 
\cite{Blake:2009jz,Drinkwater:2009sd}\footnote{The projected 
uncertainties that include reconstruction
in these references are on $H$ and $d_A$.  The uncertainty on $D$ is
inferred from these.}.  Because WiggleZ will make a 
similar measurement, such a survey would not have a dramatic
effect on cosmological parameter estimations.  However, it would
provide an excellent verification of these measurements, using
completely different methods, in different regions of the sky,
and at low cost.

Prototypes for cylindrical telescopes could perform similar science to
existing telescopes except---with dedicated resources---longer integration 
times would be feasible.  This in part makes up for the
limited resolution as for 40\,m telescopes there is a
substantial loss of information, and
only the first wiggles are resolved.  The measurements described
here would constitute a proving ground for this technique.  
The success of these prototypes would be a clear indicator of the power
and future success of full scale cylindrical telescopes.

The most powerful telescope considered is the aperture array,
which would be capable of making sub present level BAO
measurements with only a few weeks of dedicated observing.
The design for demonstrator telescope A3IV has yet to be
finalized but the proposed telescope is of nearly the same
scale as the aperture array considered here.  Depending on its
eventual configuration, the A3IV could
be a powerful probe of the Universe.

We have shown that \tcm{} intensity mapping surveys could be employed in
the short term to make useful measurements with large scale structure.
With relatively small initial resource allocations, requisite
techniques such as foreground subtraction can be tried and tested while
performing valuable science.
Such short term applications of this 
promising method will lay the trail for future dark energy
surveys.

\begin{acknowledgments}

We thank Ger~de~Bruyn for preliminary specifications of the
A3IV.
KM is supported by NSERC Canada.
PM acknowledges support of the Beatrice D. Tremaine Fellowship.

\end{acknowledgments}

\bibliography{spires,cosmo,extrarefs}

\end{document}